\documentclass{PoS}

\title{Accurate classification of 28 objects detected in the 39 months Palermo Swift/BAT hard X-ray catalogue}

\ShortTitle{Classification of 28 objects in the Swift/BAT catalogue}

\author{P. Parisi$^{1,2}$, N. Masetti$^1$, E. Jim\'enez-Bail\'on$^3$, V. Chavushyan$^4$, R. Landi$^1$, A. Malizia$^1$, E. Palazzi$^1$,
 L. Bassani$^1$, A. Bazzano$^2$, A.J. Bird$^5$, G. Galaz$^6$, D. Minniti$^{6,7}$, L. Morelli$^8$, M. Schiavone$^1$ and P. Ubertini$^2$ \\
        $^1$INAF -- IASF di Bologna, Italy.\\
        $^2$INAF -- IASF di Roma, Italy.\\
        $^3$Universidad Aut\'onoma de M\'exico, Mexico DF\\
        $^4$INAOE, Puebla, Mexico\\
        $^5$University of Southampton, UK\\
        $^6$Pontificia Universidad Cat\'olica, Santiago, Chile\\
        $^7$Specola Vaticana, Citt\`a del Vaticano\\
        $^8$Universit\`a di Padova, Italy\\
        E-mail: \email{parisi@iasf-roma.inaf.it}}

\abstract{Through an optical campaign performed at 4 telescopes 
located in the northern and the southern hemispheres, plus 
archival data from two on-line sky surveys, we have obtained 
optical spectroscopy for 28 counterparts of unclassified or 
poorly studied hard X-ray emitting objects detected with 
Swift/BAT and listed in the 39 months Palermo Swift/BAT hard 
X-ray catalogue. We have been able to pinpoint the optical counterpart of these high energy sources by means of 
X-ray observations taken with Swift/XRT
or XMM which allowed us to restrict the positional uncertainty from few arcmin to few arcsec; satellite data also provided information on the
X-ray spectra of these objects.
We find that 7 sources in our sample are Type 1 AGN while 20 are 
Type 2 AGN, with their redshifts lying between 0.009 and 0.075;
the remaining object is a Galactic cataclysmic variable (CV).
In this work we provide optical information for all  28 sources and the results of the soft X-ray analysis of
3 out of 5 AGN observed with XMM/Newton.}

\FullConference{The Extreme and Variable High Energy Sky - extremesky2011,\\
		September 19-23, 2011\\
		Chia Laguna (Cagliari), Italy}

\begin{document}

\section{Introduction}
The {\it Swift} mission was designed to study cosmic gamma-ray bursts (GRBs) in a multiwavelength context ([7]), 
but it is also able to perform dedicated X-ray and UV-optical observations
as well as surveys of the entire sky. {\it Swift} carries three instruments, i.e. the burst alert telescope (BAT; [1]), 
the X-ray telescope (XRT; [2]) and the ultraviolet/optical telescope (UVOT; [13]) and therefore 
can detect and follow up X-ray emitting objects over a wide range of wavelengths.

In particular, BAT, the high energy instrument, is a coded mask detector operating with good sensitivity in the energy range 14--195 keV 
over a 
field of view of 1.4 sr with a point source location accuracy of
 $1^{\prime}-4^{\prime}$ ([7]) depending on the source intensity.
Its sensitivity is estimated to be  $\sim$1 mCrab at high Galactic latitudes and $\sim$3 mCrab over the Galactic plane.

This instrument is not only able to detect GRBs, but also to perform  highly sensitive 
hard X-ray surveys (e.g. [4], [5], [14]). In particular, the BAT surveys allow the study of the extragalactic X-ray sky, 
and the observation of many absorbed AGNs which are often missed by lower energy instruments.
Quantifying the number of such absorbed objects, especially at low redshifts, is very important if one wants to understand the 
accretion \mbox{mechanisms} at work in AGNs and to estimate the contribution of all AGN  to the cosmic X--ray background ([3]).
However, many of the objects listed in the BAT surveys are still unclassified or poorly studied and hence they need optical follow up work to be fully characterized.

For this work we have selected from the 39 months Palermo {\it Swift}/BAT AGN survey ([4]), a group of objects (28 in total) 
either without optical identification,
or not well studied or without published optical spectra.
Following the method applied by [8] and references therein or [11] for the optical spectroscopic follow up work
of unidentified {\it INTEGRAL} and/or BAT sources, we determine the nature
of these 28 selected objects by means of X-ray observations (to pinpoint the likely X-ray counterpart) and optical measurements (to provide the source classification).

In the following sections we show the results obtained with our optical spectroscopic campaigns and we discuss in detail the results of 
the X-ray analysis of 3 out of 5 objects observed by XMM/Newton.

\section{Optical analysis}

The identification of a convincing  X-ray counterpart  of the BAT source is a fundamental step in order to reduce the positional uncertainty from 
few arcmin to few arcsec and consequently perform optical follow up work on the likely optical association/s; 
for this reason we have first analysed 
a set of X-ray observations performed using Swift/XRT for 23 out of 28 objects, while for the remaining 5 sources archival 
XMM-Newton data have been considered.
The details of the X-ray analysis can be found in [12] where we also describe the optical campaigns we have done, the telescopes used, the data reduction 
adopted as well as the classification criteria employed.
 
The optical analysis of the 28 objects in our sample reveals that 27 are AGNs and 1 is a Cataclysmic Variable (CV).
Among the sample of extragalactic objects 7  are Type 1 AGN (of which 5 are seyferts of intermediate types 1.2-1.9 and one is a 
Narrow Line seyfert 1) while 20 are 
Type 2 AGN (including a few showing LINER type signatures);  their redshifts lie between 0.009 and 0.075, i.e. they are all local AGN.
The main results of our optical study are reported in Table \ref{agn1} and \ref{agn2}, where we list for each source 
the H$_{\alpha}$, H$_{\beta}$ and [O{\sc iii}] fluxes, the classification, the redshift estimated from the narrow lines, 
the luminosity distance given in Mpc, the Galactic color excess and the color excess local to the AGN host and the name of the optical
counterpart in the NED online catalogue\footnote{\it http://nedwww.ipac.caltech.edu/}.

For the CV,  PBC J0826.3$-$7033, 
we report the H$_{\alpha}$, H$_{\beta}$  and He{\sc ii}$_{\lambda 4686}$ equivalent widths and fluxes,
the {\it R} magnitude extracted from the USNO-A2.0 catalogue ([9]), the extinction and the source distance (see Tab. \ref{cv}).

\begin{table*}[th!]
\begin{center}
\caption[]{Main results obtained from the analysis of the optical spectra of the 7 type 1 AGN of the present 
sample of {\it Swift} sources.}\label{agn1}
\scriptsize
\resizebox{12cm}{!}{
\begin{tabular}{lccccccccc}
\noalign{\smallskip}
\hline
\hline
\noalign{\smallskip}
\multicolumn{1}{c}{Object} & $F_{\rm H_\alpha}$$^*$ & $F_{\rm H_\beta}$ &
$F_{\rm [OIII]}$ & Class & $z$ & \multicolumn{1}{c}{$D_L$} & \multicolumn{2}{c}{$E(B-V)$} & \multicolumn{1}{c}{NED Name}\\
\cline{8-9}
\noalign{\smallskip}
& & & & & & (Mpc) & Gal. & AGN& \\
\noalign{\smallskip}
\hline
\noalign{\smallskip}

PBC J0503.0+2300& 699$\pm$55 &113$\pm$23& 93$\pm$7& Sy1.5 & 0.058 & 259.3 & 0.515  &0.458 & 2MASX J05025822+2259520        \\
& [2670$\pm$153] &[600$\pm$64]  & [436$\pm$45] & & & & & &\\

& & & & & & & & & \\

PBC J0543.6$-2738$  & 83.9$\pm$18.4 & 107$\pm$17 & 22.6$\pm$3.1 & Sy1.2 & 0.009 & 38.8 & 0.029 &--&ESO 424- G 012                  \\
& [88$\pm$14.9]& [120$\pm$9] & [25.4$\pm$3.2] & & & & & &\\

& & & & & & & & & \\

PBC J0814.4+0421& 394$\pm$35 & 43.9$\pm$7.7 &31.9$\pm$2.3 & NLS1 & 0.034 & 149.4 & 0.027 & 1.107& CGCG 031-072                    \\
& [433$\pm$37] & [49.5$\pm$7.8] & [36.3$\pm$5.6] & & & & & & \\

& & & & & & & & & \\

PBC J1345.4+4141 &27.3$\pm$1.5 & 0.7$\pm$0.2& 3.7$\pm$1.8 & Sy1.9 & 0.009 & 37.1 & 0.007& 2.930& NGC 5290                        \\
& [33.9$\pm$1.9] & [0.6$\pm$0.1]& [3.1$\pm$0.3] & & & & & &\\

& & & & & & & & &\\

PBC J1439.0+1413 & 16.5$\pm$3.1 & --& --& Sy1.9 & 0.072& 325.1 & 0.019&--& 2MASX J14391186+1415215         \\
& [12.9$\pm$2.9]& &  & & & & & &\\

& & & & & & & & &\\

PBC J1453.0+2553& 470$\pm$68 & 111$\pm$22 & 19.7$\pm$3.3 & Sy1 & 0.049 & 217.7 & 0.039 & 0.411 &2MASX J14530794+2554327         \\
& [1200$\pm$89] & [115$\pm$22.3] & [20.6$\pm$3.3] & & & & & &\\

& & & & & & & & &\\

PBC J1546.5+6931 & 181$\pm$20  & 26$\pm$4.5 & 97.8$\pm$8.6 & Sy1.9  & 0.037 & 162.9 & 0.041 & 1.069 & 2MASX J15462424+6929102        \\
& [244$\pm$23] & [29$\pm$4.8] & [113$\pm$16] & & & & & &\\

& & & & & & & & &\\

\noalign{\smallskip} 
\hline
\noalign{\smallskip} 
\multicolumn{10}{l}{Note: emission line fluxes are reported both as
observed and (between square brackets) corrected for the intervening Galactic} \\ 
\multicolumn{10}{l}{absorption $E(B-V)_{\rm Gal}$ along the object line of sight 
(from Schlegel et al. 1998). Line fluxes are in units of 10$^{-15}$ erg cm$^{-2}$ s$^{-1}$,} \\
\multicolumn{10}{l}{The typical error on the redshift measurement is $\pm$0.001 
but for the SDSS and 6dFGS spectra, for which an uncertainty} \\
\multicolumn{10}{l}{of $\pm$0.0003 can be assumed.} \\
\multicolumn{10}{l}{$^*$: blended with [N {\sc ii}] lines} \\
\noalign{\smallskip} 
\hline
\hline
\end{tabular} }
\end{center}
\end{table*}

\begin{table*}[th!]
\begin{center}
\caption[]{Main results obtained from the analysis of the optical spectra of the 20 type 2 AGN of the present 
sample of {\it Swift} sources.}\label{agn2}
\scriptsize
\resizebox{12cm}{!}{
\begin{tabular}{lccccccccc}
\noalign{\smallskip}
\hline
\hline
\noalign{\smallskip}
\multicolumn{1}{c}{Object} & $F_{\rm H_\alpha}$ & $F_{\rm H_\beta}$ &
$F_{\rm [OIII]}$ & Class & $z$ & \multicolumn{1}{c}{D$_L$} & \multicolumn{2}{c}{$E(B-V)$} & NED Name\\
\cline{8-9}
\noalign{\smallskip}
& & & & & & (Mpc) & Gal. & AGN & \\
\noalign{\smallskip}
\hline
\noalign{\smallskip}

PBC J0041.6+2534  & 11.3$\pm$4.5 & --& -- & Sy2/LINER & 0.015 & 65.0 & 0.035& --&NGC214 \\
& [12.1$\pm$3.8] & & & & & & & & \\

& & & & & & & & &\\

PBC J0100.6$-$4752 & 37$\pm$4 & 10.8$\pm$2.9 & 101$\pm$6 & Sy2 & 0.048 & 213.1 & 0.013 &0.251& ESO 195-IG 021           \\
& [38$\pm$4] & [10.6$\pm$3] & [106$\pm$6]  & & & & & &\\

& & & & & & & & &\\

PBC J0122.3+5004 & 139$\pm$23 & 34.1$\pm$5.7 & 169$\pm$28 & Sy2 & 0.021 & 91.4 & 0.217 &0.391& MCG +08-03-018                 \\
& [188$\pm$31]& [55$\pm$9] & [270$\pm$50] & & & & & &\\

& & & & & & & & & \\

PBC J0140.4$-$5320 & 27.3$\pm$4.3 & 4.6$\pm$0.9 &39.1$\pm$1.3 & Sy2 & 0.072 & 325.1 & 0.029 & 0.725 & 2MASX J01402676-5319389        \\
& [31.6$\pm$4.8] & [5.4$\pm$0.9] & [42.7$\pm$1.4] & & & & & &\\

& & & & & & & & &\\

PBC J0248.9+2627 & 47$\pm$4 & 4.2$\pm$0.5 & 21.6$\pm$1.5 & Sy2 & 0.057 & 274.3 & 0.158 & 1.196 &2MASX J02485937+2630391         \\
& [64.3$\pm$15.7] & [6.3$\pm$0.9] & [34.4$\pm$2.5] & & & & & &\\

& & & & & & & & &\\

PBC J0353.5+3713 & 35.5$\pm$2.8 & 6$\pm$1 & 12.8$\pm$1.3 & LINER & 0.019 & 82.6 & 0.536 & 0.938 & 2MASX J03534246+3714077        \\
& [36.7$\pm$2.8] & [5$\pm$0.9] & [13.2$\pm$1.3] & & & & & &\\

& & & & & & & & &\\

PBC J0356.9$-$4040  & 75.6$\pm$11.8& 21.7$\pm$5.6 & 167$\pm$9 & Sy2 & 0.075 & 65.0 & 0.035& 0.121&  2MASX J03565655-4041453        \\
& [70.1$\pm$7.2] &[22.4$\pm$5.6]  &[170$\pm$9]  & & & & & &\\

& & & & & & & & &\\

PBC J0544.3+5905& 3.9$\pm$0.5 & 0.7$\pm$0.1 &7.1$\pm$0.4 & Sy2 &  0.068 & 306.2 & 0.274&0.484 & 2MASX J05442257+5907361         \\
& [6.4$\pm$0.8] & [1.4$\pm$0.2] & [15.9$\pm$0.8] & & & & & &\\

& & & & & & & & &\\

PBC J0623.8$-$3212 & 97.9$\pm$17.4 &-- & 783$\pm$40 & Sy2& 0.035& 153.9 & 0.049 & --& ESO 426- G 002                 \\
&[112$\pm$13.6 ]   & & [908$\pm$47] & & & & & &\\

& & & & & & & & &\\

PBC J0641.3+3251& 51.9$\pm$6.4 & 8.2$\pm$1.7 & 197$\pm$6.2 & Sy2 & 0.049 & 217.7 & 0.153 &0.611 &2MASX J06411806+3249313         \\
& [69.4$\pm$10.4] & [13.3$\pm$2.1] & [311$\pm$20] & & & & & &\\

& & & & & & & & &\\

PBC J0759.9+2324&  8.9$\pm$1.1& 0.7$\pm$0.1 & 3.6$\pm$0.5 & Sy2  & 0.029 & 127 & 0.059 & 1.345 & CGCG 118-036                   \\
& [9.8$\pm$1.1]& [0.9$\pm$0.2] & [13.5$\pm$0.7] & & & & & &\\

& & & & & & & & &\\

PBC J0919.9+3712  &4.3$\pm$0.4  &0.5$\pm$0.1 &4.1$\pm$0.3 & Sy2 & 0.0075 & 32.3 & 0.012&1.118 & IC 2461                         \\
&[3.9$\pm$0.4] & [0.41$\pm$0.09] &3.5$\pm$0.3 & & & & & &\\

& & & & & & & & &\\

PBC J0954+3724& 8.6$\pm$0.5& 0.8$\pm$0.03 &3.6$\pm$0.6 & Sy2  & 0.019 & 82.6 & 0.016 & --& IC 2515                         \\
&[8.8$\pm$0.6]& [0.8$\pm$0.2] & [3.7$\pm$0.5] & & & & & &\\

& & & & & & & & &\\

PBC J1246.9+5433& -- & --& 16.4$\pm$4.2& Sy2 & 0.017 & 73.8 & 0.017 & --& NGC 4686                        \\
&  & & [17.8$\pm$4.4] & & & & & &\\

& & & & & & & & &\\

PBC J1335.8+0301 &19.6$\pm$2.4 & 2.9$\pm$0.5 & 16.9$\pm$1.3 & Sy2  & 0.0218 & 94.9 & 0.024 &0.830 & NGC 5231                        \\
& [20.7$\pm$1.9]& [3.2$\pm$0.5] & [17.9$\pm$1.1] & & & & & &\\

& & & & & & & & &\\

PBC J1344.2+1934 &16.6$\pm$1.8 & --& 6.6$\pm$1.1 & Sy2/LINER & 0.027 & 118 & 0.027& -- & CGCG 102-048                    \\
& [17.2$\pm$1.8] & & [6.9$\pm$1.2] & & & & & &\\

& & & & & & & & &\\

PBC J1506.6+0349 & 17.2$\pm$1.1& 2.5$\pm$0.6 & 20.8$\pm$0.8 & Sy2  & 0.038 & 167.5 & 0.049 &0.908 & 2MASX J15064412+0351444        \\
& [19.4$\pm$2.6]& [2.7$\pm$0.7]& [$23.7\pm$1.4] & & & & & &\\

& & & & & & & & &\\

PBC J2148.2$-$3455 & 6460$\pm$582& 857$\pm$83  &4970$\pm$347& Sy2  & 0.0161 & 70.7 & 0.029 & -- &NGC 7130                         \\
& [6900$\pm$895]& [947$\pm$117]&[5440$\pm$347]& & & & & &\\

& & & & & & & & &\\

PBC J2333.9$-$2343 & -- & 3.2$\pm$1.2 &14.7$\pm$2.8 & Sy2& 0.0475 & 210.8 & 0.029& -- & PKS 2331-240                    \\
& & [3.5$\pm$1.2] &[16.3$\pm$2.8 ] & & & & & &\\

& & & & & & & & &\\

PBC J2341.9+3036 & 9.7$\pm$4.6 & -- & 8.3$\pm$2.1& Sy2  & 0.017 & 73.8 & 0.102 &-- & UGC 12741                      \\
&  [14.2$\pm$5.3]& & [15.6$\pm$2.9] & & & & & &\\

& & & & & & & & &\\

\noalign{\smallskip} 
\hline
\noalign{\smallskip} 
\multicolumn{10}{l}{Note: emission line fluxes are reported both as 
observed and (between square brackets) corrected for the intervening Galactic} \\ 
\multicolumn{10}{l}{absorption $E(B-V)_{\rm Gal}$ along the object line of sight 
(from Schlegel et al. 1998). Line fluxes are in units of 10$^{-15}$ erg cm$^{-2}$ s$^{-1}$,} \\
\multicolumn{10}{l}{The typical error on the redshift measurement is $\pm$0.001 
but for the SDSS and 6dFGS spectra, for which an uncertainty} \\
\multicolumn{10}{l}{of $\pm$0.0003 can be assumed.} \\
\noalign{\smallskip} 
\hline
\hline
\end{tabular}}
\end{center}
\end{table*}

\begin{table*}[th!]
\caption[]{Main results concerning PBC J0826.3$-$7033
identified as a cataclismic variable.}\label{cv}
\hspace{-1.2cm}
\scriptsize
\vspace{-.5cm}
\begin{center}
\resizebox{13cm}{!}{
\begin{tabular}{lcccccccccr}
\noalign{\smallskip}
\hline
\hline
\noalign{\smallskip}
\multicolumn{1}{c}{Object} & \multicolumn{2}{c}{H$_\alpha$} & 
\multicolumn{2}{c}{H$_\beta$} &
\multicolumn{2}{c}{He {\sc ii} $\lambda$4686} &
Optical & $A_V$ & $d$ & \multicolumn{1}{c}{$L_{\rm X}$} \\
\cline{2-7}
\noalign{\smallskip} 
 & EW & Flux & EW & Flux & EW & Flux & mag. & (mag) & (pc) & \\
\noalign{\smallskip}
\hline
\noalign{\smallskip}
PBC J0826.3$-$7033 & 38.9$\pm$1.8 & 66$\pm$3 & 33.6$\pm$1.5 & 44$\pm$2 & 5.7$\pm$0.9 & 7.4$\pm$1.1&
 13.8 ($R$) & 0 & 90 & 0.002 (2--10) \\
 & & & & & & & & & & 0.007(20--100) \\
\noalign{\smallskip} 
\hline
\noalign{\smallskip}
\hline
\multicolumn{11}{l}{Note: EWs are expressed in \AA, line fluxes are
in units of 10$^{-15}$ erg cm$^{-2}$ s$^{-1}$, whereas X--ray luminosities
are in units of 10$^{33}$ erg s$^{-1}$} \\
\hline
\hline
\end{tabular}}
\end{center} 
\end{table*}

\section{XMM analysis}

We report the results obtained from the X-ray data analysis of 3 out of 5 sources observed with XMM/Newton. We used data acquired with the pn X-ray CCD camera on
the EPIC instrument on-board XMM. The other two objects were excluded for the following reasons:
PBC J0041.6+2534 has a  very low quality X-rays spectrum while  the XMM data of  PBC J0919.9+3712 have  already  been reported in the literature by Noguchi et al. (2009).

These data were processed using the Standard Analysis Software (SAS) version 9.0.0
employing the latest available calibration files and following usual procedures described in details in [12].
The spectral analysis has been performed using 
{\sc XSPEC} v.12.6.0 and assuming initially a simple power law passing  through Galactic  ([6]) and intrinsic absorption;
if this baseline model was not sufficient to fit the data, we then introduced extra spectral components as required, 
according to the F-test statistics.

The results of this analysis are reported in 
Table \ref{X}, where we list the Galactic absorption, the column density in excess to the Galactic value, 
the power law photon index, the reduced 
$\chi^{2}$ of the best-fit model, the 2--10 keV flux and 20--100 keV fluxes; extra spectral parameters if required are discussed in the text.
Quoted errors correspond to 90$\%$ confidence level for a single
parameter of interest ($\Delta\chi^{2}=2.71$).

\subsection{PBC J1246.9+5433}
The best fit model 
 ({\it wa$_{gal}$*(bb+wa*(po+ga+ga))}) to this source 
requires an extra black body component with a $kT = 0.28^{+0.04}_{-0.03}$ keV 
and two narrow lines at $E = 6.29^{+0.03}_{-0.03}$ keV
and $6.79^{+0.11}_{-0.10}$ keV with an EW  of $600^{+182}_{-174}$ eV and $378^{+168}_{-167}$ eV respectively.
The other parameters of the baseline model are reported in Table 4.
The presence of  strong  excess emission  below 1 keV and of two prominent lines around 6.3 and 6.8 keV, 
together with the extremely flat power law strongly point to a highly absorbed AGN. Indeed, PBC J1246.9+5433 shows an 
intrinsic absorption of 23.6$^{+8.0}_{-9.6}$ $\times$ 10$^{22}$ cm$^{-2}$. 
This object has been classified as a Seyfert 2  from our optical spectroscopic analysis and its X-ray spectrum 
is fully compatible with its optical type.

\subsection{ PBC J1335.8+0301}
The best $\,$fit $\,$ for$\,$ the X-ray $\,$spectrum $\,$of this$\,$  source $\,$is obtained$\,$using a$\,$ simpler $\,$ model
({\it wa$_{gal}$*(po+wa*po)}) than that of PBC J1246.9+5433. With respect to our baseline model, we only found  
an  extra power law component having the same photon index of the primary absorbed power law but
passing only through the Galactic column density. The photon index is flat ($\Gamma=$ 1.58$^{+0.06}_{-0.04}$) while the intrinsic 
column density is 2.3$^{\pm 0.1}$ $\times$ 10$^{22}$ cm$^{-2}$. Also PBC J1335.8+0301 is classified as a Seyfert 2 galaxy in optical, 
although  the X-ray data suggest that this is a mildly absorbed type 2 AGN.

\subsection{PBC J2341.9+3036}
Also in this case the  best fit model shows some extra features ({\it wa$_{gal}$*(po+wa*(po+ga))}). It requires  
a second power law component
having the same photon index of the primary absorbed power law 
and passing only through the Galactic column density, as well as   
a narrow emission line at $E = 6.25^{+0.05}_{-0.06}$ with an EW $= 365^{+176}_{-147}$ eV.
The primary continuum has a steep photon index ($\Gamma =$ 2.02$^{+0.16}_{-0.15}$) and a column density of 
56.5$^{+15.5}_{-10.4}$ $\times$ $10^{22}$ cm$^{-2}$, which  
makes PBC J2341.9+3036 one of the most absorbed sources in our sample, in agreement with its optical classification as a Seyfert 2 galaxy.

\begin{table*}[th!]
\begin{center}
\caption[]{Main results obtained from the analysis of the X-ray spectra of 3 out of 5 objects observed with XMM/Newton in the present 
sample.}\label{X}
\scriptsize
\resizebox{13cm}{!}{
\begin{tabular}{lcccccc}
\noalign{\smallskip}
\hline
\hline
\noalign{\smallskip}
\multicolumn{1}{c}{Source} & \multicolumn{1}{c}{N$_{Hgal}$}  & $\Gamma$ & N$_H$  &  $\chi^2/\nu$   & F$_{(2-10)keV}$ &F$_{(20-100)keV}$ \\
\noalign{\smallskip}
& $\times 10^{22}$cm$^{-2}$& &$\times 10^{22}$cm$^{-2}$ & &$\times 10^{-11}$erg s$^{-1}$ cm$^{-2}$&$\times 10^{-11}$erg s$^{-1}$ cm$^{-2}$ \\
\noalign{\smallskip}
\hline
\noalign{\smallskip}
& & & & & & \\

 PBC J1246.9+5433 &0.014 &  0.88$^{+0.13}_{-0.12}$ & 23.6$^{+8.0}_{-9.6}$  & 15/22 &0.09 & 1.5 \\
& & & & & &\\

 PBC J1335.8+030 & 0.019 & 1.58$^{+0.06}_{-0.04}$ & 2.3$^{+0.1}_{-0.1}$ &313.3/299 & 0.64 & 1.1  \\
& & & & & &\\

 PBC J2341.9+3036 &0.058  & 2.02$^{+0.16}_{-0.15}$ & 56.5$^{+15.5}_{-10.4}$ & 25/28 & 0.11 & 1.0 \\
& & & & & &\\

\hline
\hline
\end{tabular}}
\end{center}
\end{table*}

\section{Conclusions}
With this work we have been able to either give or confirm or correct  the optical classification of 28 {\it Swift} sources 
belonging to the Palermo 39 months {\it Swift} catalogue (see also [12] for details). 
This was achieved through a multisite observational campaign in Europe, South Africa, Central and South America.

We found that our sample is composed of 27 AGNs (7 of Type 1 and 20 of Type 2), with redshifts between 0.009 and 0.075, 
and 1 CV.
Among these sources we found some peculiar objects, such as 3 likely LINERs and 1 narrow line seyfert 1.

The X-ray spectral analysis of 3 out of 5 sources observed with XMM-Newton shows a complex best-fit model with an absorbed power law component as a primary emission model; all 3 require an extra component 
at low energies to fit an emission excess below few keV, while only two display iron line emission features.

This work shows the importance of optical spectroscopic follow up observations for sources discovered by hard X-ray  surveys 
and either unclassified or poorly studied.
By increasing the number of the identifications in hard X-ray catalogues, it is possible to perform more reliable statistical studies as  
multiwavelength characterization of the sources, thus allowing a better understanding of  the physical processes that drive the powerful AGN detected.

\end{document}